\documentstyle[12pt,aaspp4,flushrt]{article}

\begin{document}

\title{The Mass distribution of the Cluster 0957+561 from Gravitational
Lensing}

\author{Philippe Fischer\footnote{Visiting Astronomer, Canada France Hawaii
Telescope}\footnote{Hubble Fellow} and Gary Bernstein$^{1}$
}
\affil{Dept. of Astronomy, University of Michigan, Ann Arbor, MI 48109}

\author{George Rhee$^{1}$}

\affil{Dept. of Physics, University of Nevada Las Vegas, Las Vegas, NV 89195}

\author{ J. Anthony Tyson}
\affil{Bell Laboratories, 600 Mountain Ave., Murray Hill, NJ 07974} 

\abstract

Multiply gravitationally lensed objects with known time delays can lead to
direct determinations of H$_0$ independent of the distance ladder if the mass
distribution of the lens is known. Currently, the double QSO 0957+561 is the
only lensed object with a precisely known time delay. The largest remaining
source of systematic error in the H$_0$ determination results from uncertainty
in the mass distribution of the lens which is comprised of a massive galaxy
(G1) and the cluster in which it resides.

We have obtained V-band CCD images from CFHT in order to measure the mass
distribution in the cluster from its gravitional distorting effect on the
appearance of background galaxes. We use this data to constuct a
two-dimensional mass map of the field. A mass peak is detected at the
$4.5\sigma$ level, offset from, but consistent with, the position of G1.
Simple tests reveal no significant substructure and the mass distribution is
consistent with a spherical cluster. The peak in the number density map of
bright galaxies is offset from G1 similarly to the mass peak.

We constructed an azimuthally averaged mass profile centered on G1 out to
2\arcmin\ ($400 h^{-1}$ kpc). It is consistent with an isothermal mass
distribution with a small core ($r_c \approx 5\arcsec\ = 17 h^{-1}$ kpc). The
inferred mass within 1 Mpc is consistent with the dynamical mass estimate but
$2\sigma$ higher than the upper limits from a ROSAT X-ray study.

We discuss implications for H$_0$ in a future paper.

\section{Introduction}

One of the most interesting cosmological applications of gravitational lensing
is the determination of Hubble's constant. This can be achieved by measuring
the time delay between the observed light curves of multiply imaged QSOs
(Refsdal, 1964). If the redshifts of the lens and source are known, and the
mass distribution of the lens is well-understood, H$_0$ can be determined
directly. Two advantages of this technique are that it is independent of the
distance ladder-based methods and it is insensitive to motions of the nearby
Universe.

%

The first-discovered doubly imaged quasar Q0957+561 (Walsh et al. 1979) has
been the subject of intensive monitoring campaigns and is currently the only
lensed system with a well-determined time delay (Thomson \& Schild 1994, Schild
\& Thomson 1995, Kundic et al. 1996). The lens is comprised of a primary lens (the
galaxy G1) and the cluster containing G1 (Young et al. 1981, Bernstein et
al. 1993, Grogin \& Narayan 1996). The largest remaining uncertainty in H$_0$
derived from this system is due to the uncertainty in the mass distribution in
the lens, specifically, the ratio of the galaxy surface mass density to the
cluster surface mass density.

An attempt has been made to measure the mass of G1 spectroscopically, yielding
a velocity dispersion of $300 \pm 50$ km s$^{-1}$ (Rhee 1991). However, this is
a difficult measurement due to the proximity of the QSO B image. The
spectroscopic value is somewhat lower than the $\sim 400$ km s$^{-1}$ implied
by the Faber-Jackson law (Bernstein et al. 1993).

In this paper we describe the direct measurement of the 0957+561 cluster
surface mass density by studying its distorting effect on the appearance of
background galaxies. This technique has recently been demonstrated to work well
on massive clusters (e.g. Tyson et al. 1990, Tyson \& Fischer 1995, Squires et
al 1996), however, the cluster in the 0957+561 field is believed to be of
significantly lower mass than previously studied clusters and will be a
difficult test of the method.

In \S \ref{observations} and \S \ref{profit} we describe the observations and
analysis. \S \ref{reconstruct} describes the mass reconstruction techniques.
\S \ref{calibration} describes simulations which were carried out in order to
calibrate the mass measurements and quantify the uncertainties and systematic
errors in the analysis. \S \ref{discuss} discusses the 2-d mass maps and
azimuthally averaged radial mass profile, \S \ref{other} compares the lensing
results to previous mass determinations and \S \ref{conclusion} contains the
conclusions and a description of some future work.

\section{Observations} \label{observations}

Twenty-nine 720s V-band images were taken of the region around the double
quasar Q0957+561 using the Canada France Hawaii Telescope on 10-11 Jan
1994. FOCAM was used with the thick LORAL3 2048$^2$ CCD. This chip has
0.207\arcsec\ pixels and a field size 6\arcmin\ on a side. The effective field
size is reduced somewhat because of vignetting at the corners due to the filter
wheel. The conditions were photometric. The telescope was dithered between
observations which allowed us to construct a flat-field from our observations.
The final combined V-band image has a useable field size of 5.8\arcmin\
$\times$ 5.5\arcmin\ with FWHM = 0.6\arcsec\ and the RMS noise of the sky is
approximately 28.7 V mag per square arcsec (Fig. \ref{fixV}).

\begin{figure}
\caption{V-band CFHT image of the field centered at
0957+561. The image is $5.8\arcmin\ \times 5.5\arcmin$, north is at the top and
east is to the left.  The total exposure time is 5.8 hours. The seeing is FWHM
= 0.6\arcsec\ and the RMS noise of the sky is approximately 28.7 V mag per
square arcsec. The image intensity scaling is logarithmic. \label{fixV}} 
\end{figure}

The photometric zeropoint for the 0957+561 V image was determined to an
estimated accuracy of 0.03 mag from observations of five standard stars in the
NGC 2264 field (Christian et al. 1985). Reddening in the 0957+561 field is low,
E(B--V) $<$ 0.03 mag (Burstein \& Heiles 1982) so we assume E(B--V) = 0 in this
paper.

We located 21 bright isolated stars with V $<$ 23 to investigate the behaviour
of the point-spread-function (PSF). These had measured ellipticities spanning
$0.00 \le \epsilon \le 0.03$ ($\epsilon$ = 1-b/a) with a mean of $\epsilon =
0.015$. The 20 objects with $\epsilon > 0$ have position angles ranging from
$-55 < $PA (deg) $< 10$, implying a very slight position independent trailing,
possibly arising from guiding errors or during the image combining
procedure. This effect must be very small since any uncertainty in the
ellipticity measurements of round objects will tend to result in an
overestimate in the ellipticity on average. We test systematics arising from
this PSF anisotropy by carrying out simulations using the PSF derived from the
image (see \S \ref{calibration}).

\section{Faint Galaxy Photometry and Analysis} \label{profit}

The faint galaxy analysis was carried out using the analysis software ProFit
(developed by PF). This software, starting with the brightest objects, fits an
analytical model to each galaxy, using weighted, non-linear least squares, and
subtracts the galaxy light from the image. It then proceeds to successively
fainter objects. Once it has detected and subtracted all the objects in an
image it replaces each galaxy in turn and refits and resubtracts until
convergence is achieved. The software outputs brightness, orientation,
ellipticity and other image parameters based on the fitted function. Fig
\ref{fixVs} shows the final subtracted image.

\begin{figure}
\caption{The same as Fig. \protect\ref{fixV} after ProFit
galaxy fitting and subtraction. \label{fixVs}}
\end{figure}

\begin{figure}
\plotone{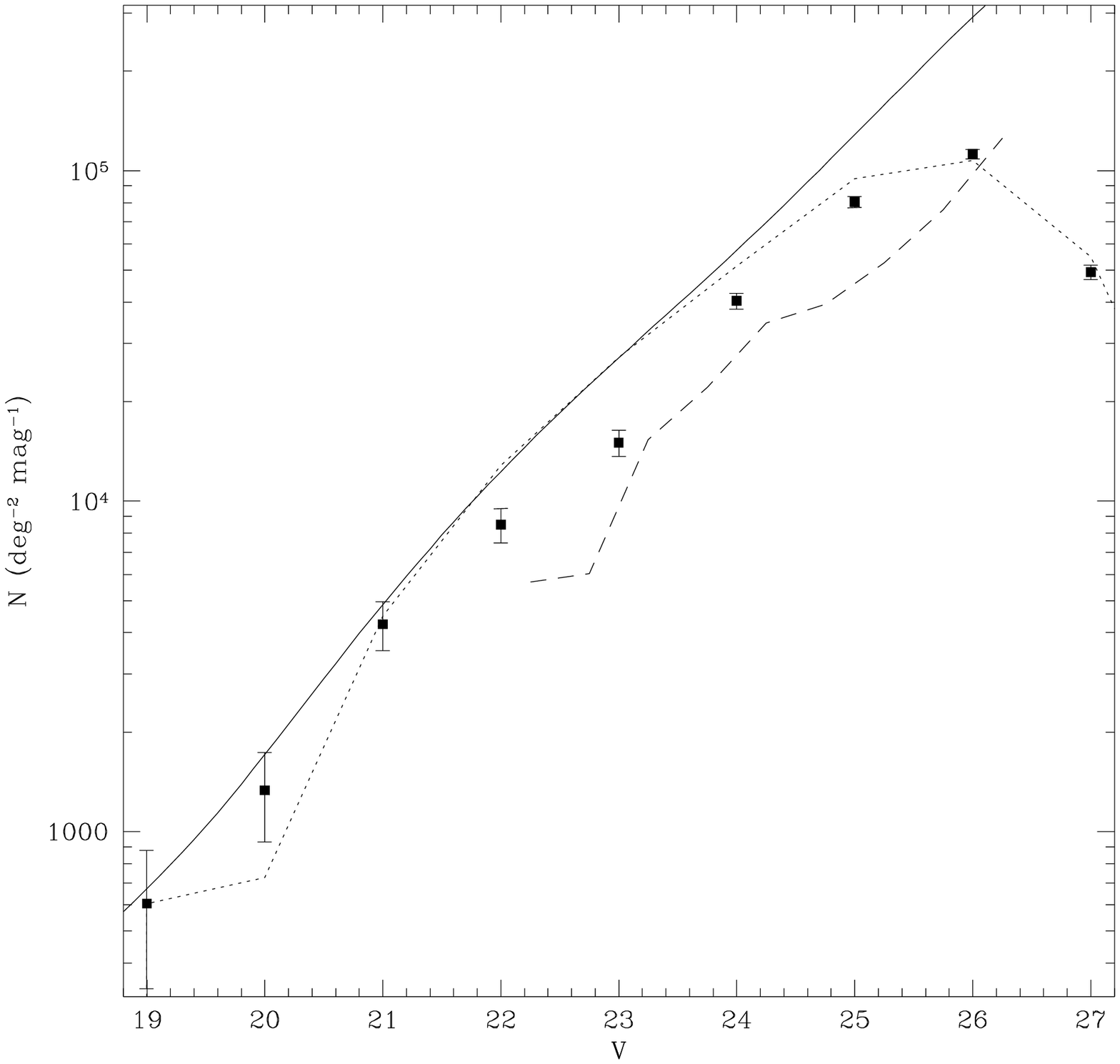}
\caption{Galaxy plus star counts for the observed image
(points) and one of the simulations (dotted line). The galaxy magnitudes are
isophotal magnitudes with outer isophote of 27.8 mag square arcsecond.  The
error bars are based on Poisson statistics. The solid line shows the input
function for stars and galaxies. Also shown are recent V-band counts (dashed
line) from Smail et al. (1995) which are about 45\% below or half a magnitude
to right of the 0957+561 counts. \label{lumfunc}}
\end{figure}

\begin{figure}
\plotone{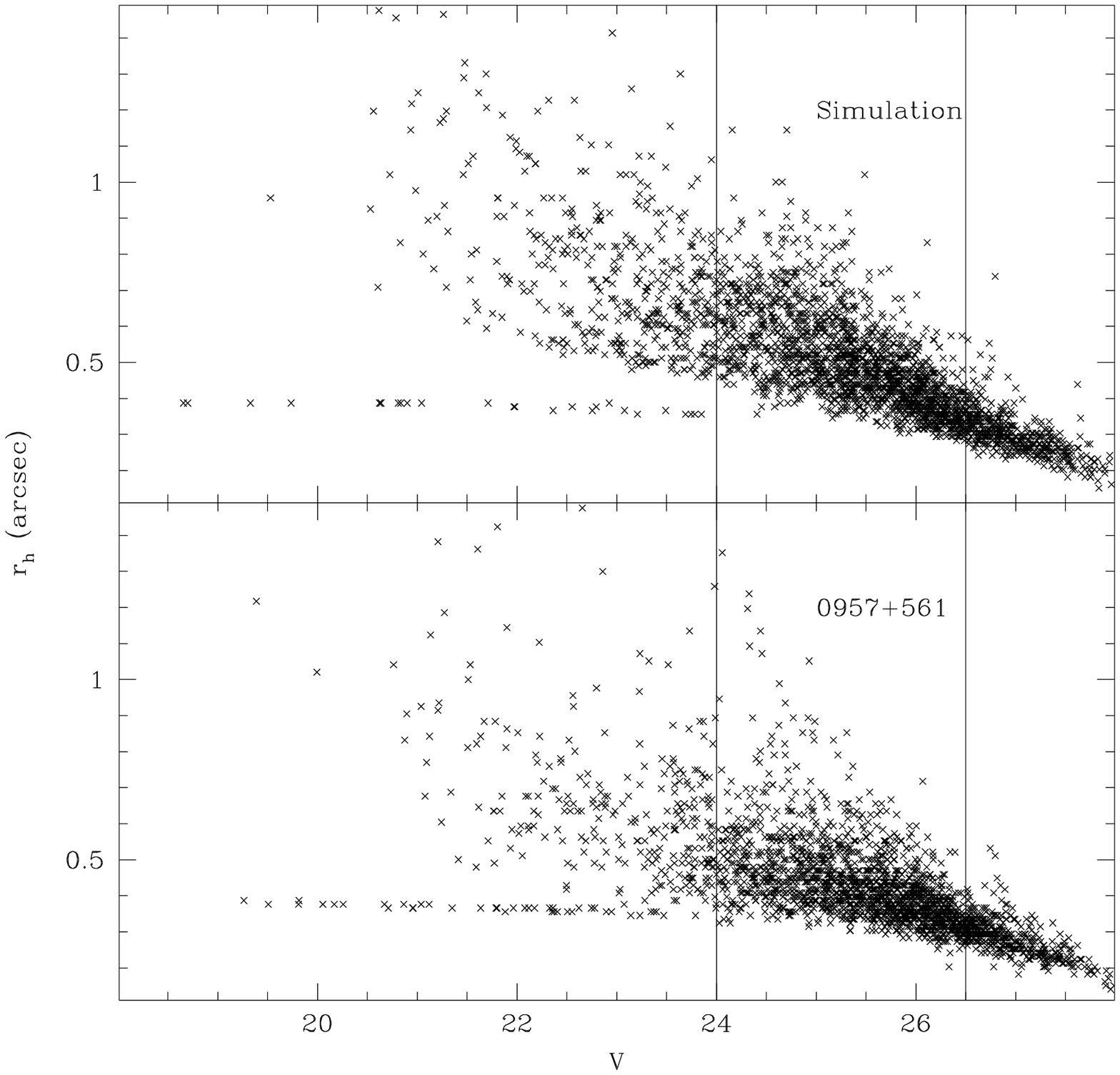}
\caption{Half light radius vs. apparent V magnitude for
the data (lower panel) and one of the Monte-Carlo simulations (upper
panel). The radii are not corrected for seeing. Good quantitative agreement is
seen. The vertical lines indicate the magnitude range of galaxies used in the
mass reconstruction \label{sizemag}}
\end{figure}

Fig. \ref{lumfunc} shows galaxy counts for the faint galaxies in the field of
Q0957+561. The magnitudes are isophotal magnitudes with outer isophote of 27.8
mag square arcsecond (31.2 mag per pixel). Also shown are recent V-band blank
field counts from Smail et al. (1995); while the slopes of the counts are in
good agreement, the Smail et al. counts lie about 45\% below or 0.5 magnitudes
to the right of our counts. Fig. \ref{sizemag} shows the size-magnitude
relationship for the faint galaxies ($r_h$ is the radius which contains half
the light, uncorrected for seeing). The galaxy sequence is well-separated from
the stellar sequence down to about 24th mag.

\section{Mass Reconstruction} \label{reconstruct}

For gravitational lensing, the relationship between the tangential shear,
$\gamma_T$, and surface mass density, $\Sigma$, is (Miralda-Escud\'e 1991,
1995),

\begin{equation}\label{escude}
\gamma_T(r) = \overline{\kappa}(\le r) - \overline{\kappa}(r),
\end{equation}

\noindent
where $\kappa = \Sigma/\Sigma_{crit}$ and $r$ is the radial distance from a
given point in the mass distribution. The first term on the right is the mean
density interior to $r$ and the second term is the mean density at $r$. The
basic technique for 2-d surface mass density reconstruction using the
distortions of faint background galaxies is outlined in Kaiser \& Squires
(1993) (the KS algorithm). For a given coordinate on the image ($\vec{R}$), the
distortion quantity for the i$^{th}$ galaxy is:

\begin{equation} 
D_i(\vec{R}) = {1-(b_i/a_i)^2 \over 1 + (b_i/a_i)^2} \times
{[\cos(2\theta_i)(\Delta{x_i}^2 - \Delta{y_i}^2) +
2\sin(2\theta_i)\Delta{x_i}\Delta{y_i}] \over \Delta{x_i}^2 + \Delta{y_i}^2},
\end{equation}

\noindent where $(b_i/a_i)$ and $\theta_i$ are the galaxy axis ratio and
position angle, respectively. $\Delta x$ and $\Delta y$ are the angular
horizontal and vertical distances from $\vec{R}$ to galaxy $i$, and $D$ is
related to the tangential shear by (Schneider \& Seitz 1995):

\begin{equation}
<D(r)> = 2{\gamma_T(r)[1-\kappa(r)] \over [1-\kappa(r)]^2 + \gamma_T^2(r)}
\end{equation}

\noindent
In the weak lensing regime $\kappa << 1$, and $\gamma_T << 1$, $\gamma_T
\approx <D>$ and the surface mass density is given by:

\begin{equation}\label{kseqn}
\kappa(\vec{R}) = {1\over \overline{n}\pi}\sum_{i=1}^{N}{D_i(\vec{R})\over
\Delta{x_i}^2 + \Delta{y_i}^2},
\end{equation}

\noindent 
where $N$ is the number of galaxies and $\overline{n}$ is the number density of
galaxies. Eqn. \ref{kseqn} assumes that the galaxies are intrinsically (in the
absence of lensing) randomly aligned.  The critical density, $\Sigma_{crit}$
depends on the redshift distribution of the background galaxies. Because of the
random intrinsic alignments of the background galaxies, the formal error of
$\kappa$ from Eqn. \ref{kseqn} is infinite and in practice one must employ a
smoothing kernel:

\begin{equation}
\kappa(\vec{R}) = {1\over
\overline{n}\pi}\sum_{i=1}^{N}{W(\Delta{x},\Delta{y},s)D_i(\vec{R})\over
\Delta{x_i}^2 + \Delta{y_i}^2}
\end{equation}

In this paper we have have used a smoothing kernel of the form (Seitz \&
Schneider 1995):

\begin{equation}
W(\Delta{x},\Delta{y},s)=1-\left(1+{\Delta{x}^2+\Delta{y}^2 \over
2s^2}\right)e^{-(\Delta{x}^2+\Delta{y}^2)/2s^2},
\end{equation}

\noindent
where `$s$' is referred to as the smoothing scale.

The 2-d mass map for the 0957+561 field is shown in Fig. \ref{contour}. We
discuss this further in \S \ref{2d}.

\begin{figure}
\caption{Mass map of the 0957+561 field using the
technique of Kaiser \& Squires (1993) superimposed on a V-band CCD image. A
total of 1651 galaxies (arclets) with $24.0 \le V \le 26.5$ were used in the
reconstruction and the the smoothing scale is $s=30\arcsec$. The contours are
stepped in units of $0.5\sigma$. North is up and East is to the left. The
displayed field is 327\arcsec\ on a side. \label{contour}}
\end{figure}

Because of the smoothing kernel, plus biases introduced by edge effects in the
images, this formula is mainly useful for determining the 2-d shapes of mass
distributions. A less biased way of obtaining mass estimates as well as
azimuthally averaged density profiles is (Fahlman et al. 1994):

\begin{eqnarray}\label{denseqn}
\overline\kappa(r \le r_i) - \overline\kappa(r_i \le r \le r_o) & = &
{1\over 2\pi\overline{m}({1 - r_i^2/r_o^2})}\sum_{r_i \le r \le
r_o}{D_i(\vec{R})\over \Delta{x_i}^2 + \Delta{y_i}^2} \nonumber\\ & & \\
& = & {r_o^2 \over 2N_{io}}\sum_{r_i \le r \le
r_o}{D_i(\vec{R})\over \Delta{x_i}^2 + \Delta{y_i}^2}, \nonumber
\end{eqnarray}

\noindent
where $N_{io}$ is the number of galaxies between $r_i$ and $r_0$, and
$\overline{m}$ is the corresponding number density.

\begin{figure}
\plotone{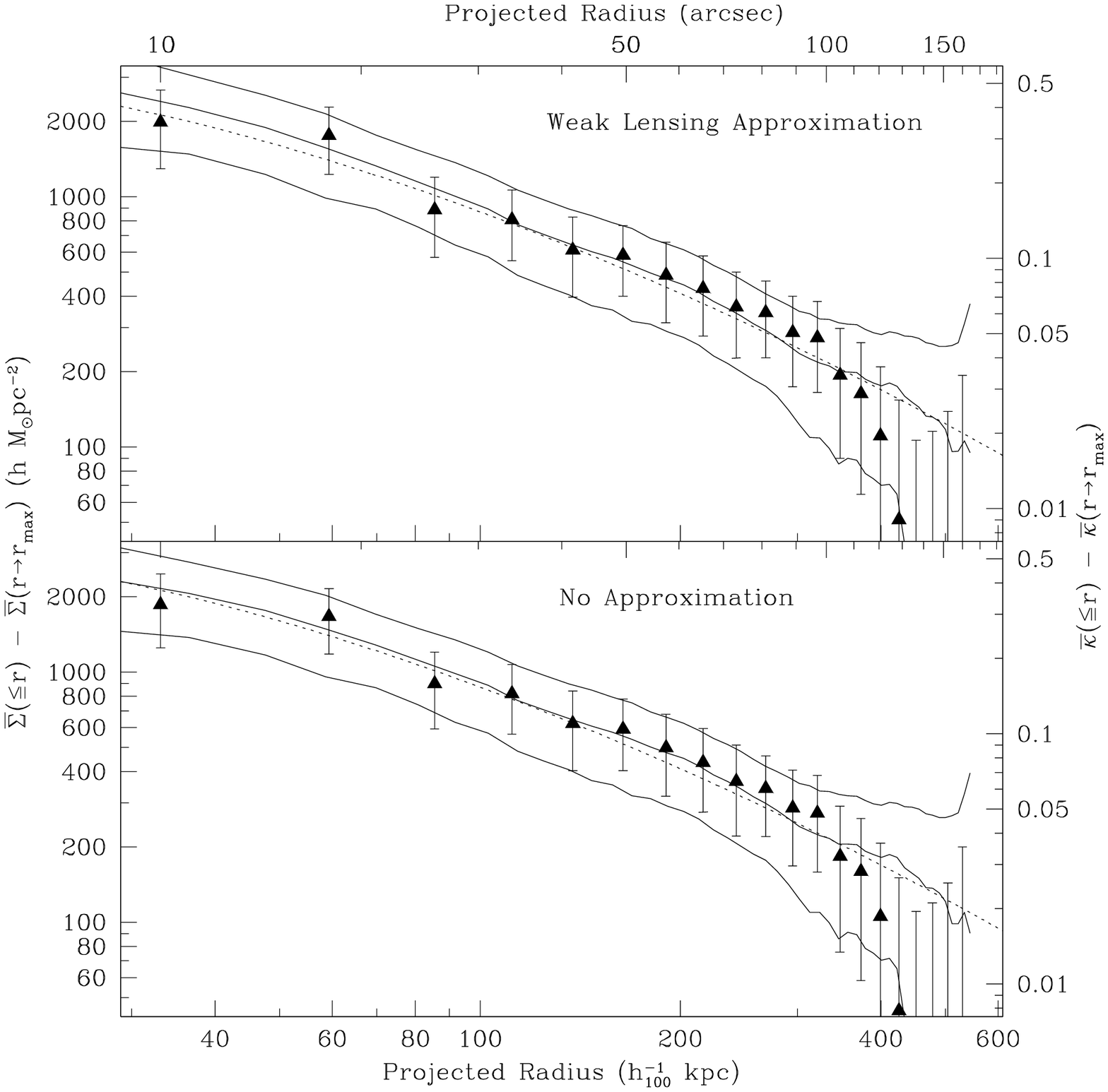}
\caption{The upper panel is the radial mass density profile
for the 0957+561 cluster from Eqn. \protect\ref{denseqn} (weak lensing
approximation) centered on G1. The points are the data for the cluster derived
from 1307 galaxies having $24.0 \le V \le 26.5$.  The solid line is the mean
profile and the 1$\sigma$ upper and lower limits for 102 simulations of a
spherical cluster having $r_c = 5\arcsec\ = 17 h^{-1}$ kpc and $\Sigma_0 = 3600
h$M$_\odot$ pc$^{-2}$. $r_{max} = 168\arcsec\ = 550 h^{-1}$ kpc. The input
model is shown by the dotted line. The lower panel is the radial density
profile from Eqn. \protect\ref{nonweakeqn}. Both the data and the simulations
have been corrected using the values of $\kappa$ and $\gamma_T$ from the input
profile for the simulations (see \S \protect\ref{profile}). \label{compdens}}
\end{figure}

In the upper panel of Fig. \ref{compdens} we show the calibrated 0957+561
cluster radial mass density profile resulting from Eqn. \ref{denseqn}. The
error bars are based on the 1$\sigma$ galaxy-to-galaxy scatter of the values
inside the sum in Eqn. \ref{denseqn}. Details of the mass calibration and
systematic errors can be found in \S \ref{calibration} and a discussion
of the radial mass profile is in \S \ref{profile}.

It should be mentioned that galaxy distortion is insensitive to flat sheets of
mass. Consequently, all mass measurements described in this paper are uncertain
by an unknown additive constant. If there is a substantial flat component to
the mass distribution our mass estimates will be lower limits (see \S
\ref{profile}).

\section{Calibration and Systematics} \label{calibration}

Application of the mass reconstruction techniques described above will result
in underestimates of the mass due to: measurement error, the effects of seeing,
and the intrinsic ellipticity of the background galaxies. These effects must be
calibrated in order to correct the mass estimates. Calibration was carried out
using two methods. The first involved applying a known constant shear to the
Hubble Deep Field (HDF) (Williams et al. 1995) and the second was a full
Monte-Carlo simulation of the data.

\subsection{Hubble Deep Field Simulations} \label{hdf}

Compared to the CFHT data the HDF data is both much higher resolution and much
deeper. Therefore, it offers an excellent opportunity to calibrate our mass
estimates as described in Kaiser et al. (1995). The technique involves
stretching the HDF data by $1+\delta$, convolving with the PSF and adding
noise. The values of $D_i$ are measured for each galaxy and compared to the
unstretched values of $D_i$. The quantity of interest is the recovery factor,
$C = <\Delta{D_i}>/\delta$. The raw values of $\kappa$ are then multiplied by
$1/<C>$ to yield the corrected values. One problem is that the HDF V-band
filter (F606W) is significantly different from our V-band filter in both
central wavelength and width. In order to insure that we were sampling the same
population of galaxies we chose a 2.5 magnitude range in the HDF data which
yielded the same surface number density of galaxies as in our field for the
range $24.0 \le V \le 26.5$ (see \S \ref{2d}). This turned out to be $23.7 \le
F606W_{HDF} \le 26.2$. This V-F606W magnitude difference is roughly equivalent
to the range of values calculated by Fukugita et al. (1995) for late-type
galaxies at intermediate to high redshift. We stretched the HDF data in two
different directions and obtained a value of $<C> = 0.37 \pm 0.03$.

\subsection{Monte-Carlo Simulations} \label{monte}

The HDF simulations yield $<C>$ which is appropriate for a constant distortion
of all galaxies. In reality galaxies at different redshifts are distorted by
different amounts (i.e. high-$z$ galaxies are more distorted). Therefore, if
galaxy size in a fixed apparent magnitude range is a function of redshift, one
would expect $C$ to vary as a function of redshift resulting in a mean value
which differs from that of the HDF simulations.  By carrying out full 3-d
Monte-Carlo simulations of the observed data we can quantify this difference.
Aside from calibrating the lensing results, simulations of the observational
data are useful for quantifying systematics in the inferred mass distributions
and determining $\Sigma_{crit}$.

The simulations consist of artificial galaxies distributed in seven redshift
shells (z = 0.0 - 0.355, 0.355 - 0.5, 0.5 - 0.65, 0.65 - 0.8, 0.8 - 1., 1.0 -
2.0, 2.0 - 7.2). For each shell, simulated images are made with pixels equal to
one quarter the area of the pixels on the observed image. Galaxy images are
generated for each shell based on the quiescent models of McLeod \& Rieke
(1995). Briefly, the McLeod \& Rieke models involve six galaxy types (E, S0,
Sab, Sc, Sd, Im) with a formation epoch of z = 7.2. H$_0 = 60$ km s$^{-1}$
Mpc$^{-1}$ and q$_0 = 0.05$ were assumed. Evolution and k-corrections are based
on the synthesis models of Bruzual \& Charlot (1993). Bulge-to-disk ratios are
taken from King \& Ellis (1985). The model counts were normalized to fit our
V-band galaxy counts. Stars were also added to the simulated images based on
the observed star counts (measured for $V < 24$ and extrapolated with the same
slope to the detection limit).

Initially we used galaxy size - absolute magnitude relationships from Im et
al. (1995), however, these were seen to produce apparent magnitude - size
relationships substantially larger than observed for V $> 24$. Smail et
al. (1995) also saw small galaxy sizes for V $> 24$ with similar image quality.
The evolution of galaxy sizes (as well as other aspects of galaxy evolution) is
not well understood. One explanation for the small galaxies we observe is that
we are seeing more small galaxies at low to intermediate redshift than
predicted by the McLeod \& Rieke models. This might also explain why these
models underpredict the density of apparently faint galaxies. Alternatively,
higher redshift galaxies may have their emission more concentrated towards
their centers than local galaxies or we may be seeing small emitting regions
within the galaxies. It is not currently possible to distinguish between these
(and other) scenarios. For the present work, we arbitrarily assume that the
galaxy sizes evolve as (1+z)$^{-n}$. We determine $n$ by stretching the
simulated data by a constant amount as was done for the HDF simulations and
modifying $n$ until we get the same value of $<C>$ as is observed for the HDF
data in the range $24.0 \le $ V $ \le 26.5$. This yielded $n = 0.75$; the
size-apparent magnitude relation for the simulated galaxies are in good
agreement with the faint galaxy measurements down to the detection limit (see
Fig. \ref{sizemag}).

The galaxies in each shell were distorted with various spherically symmetric
mass distributions located at z=0.355 under the assumption that all galaxies
within a shell lie at the number weighted mean redshift of that shell. The
distorted images were combined and convolved with the slightly elongated PSF
derived from the observations. The images were block-summed $2\times2$ and
Gaussian noise was added. The simulated images were processed identically to
the observed images.

\begin{figure}
\plotone{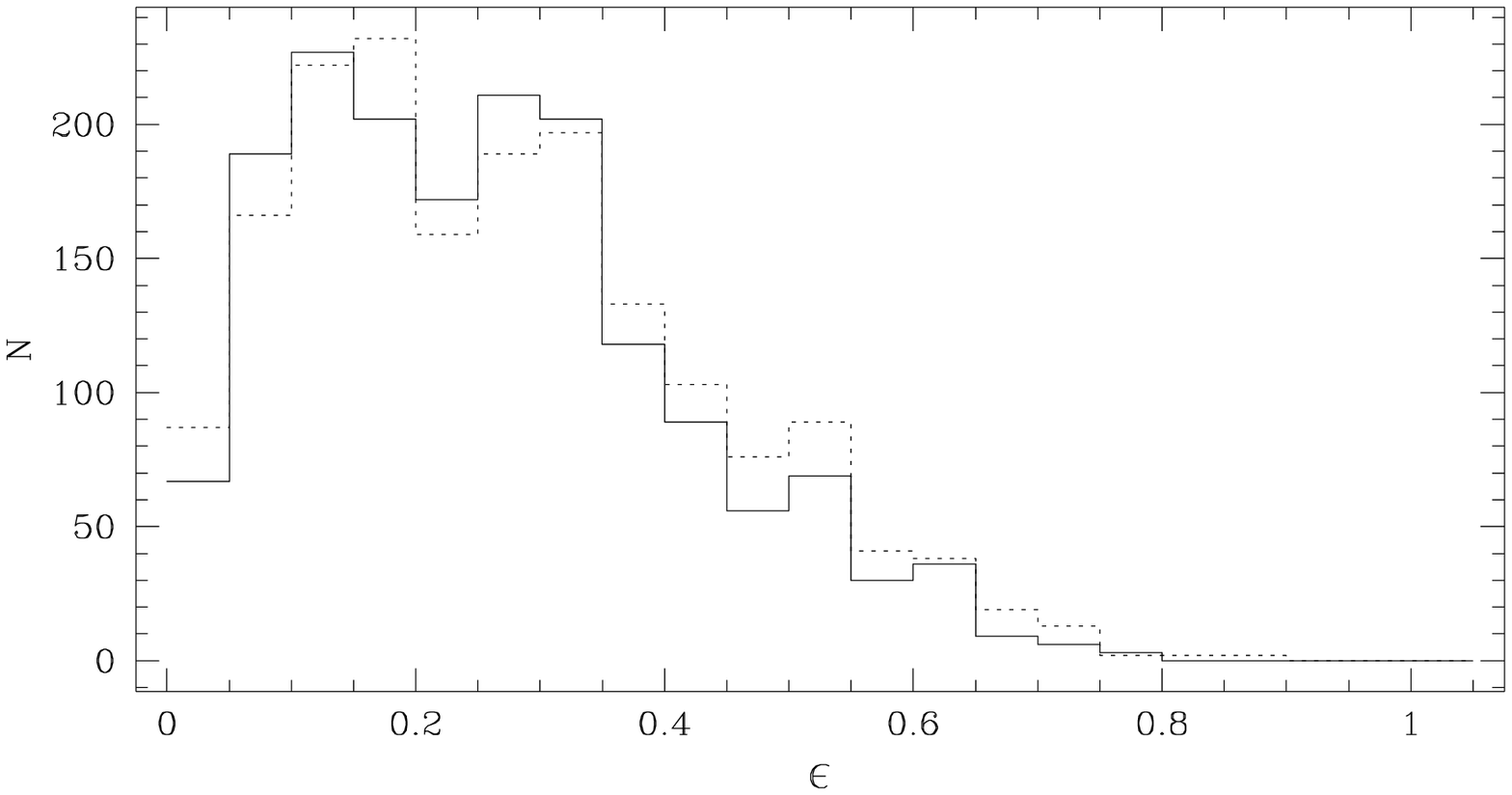}
\caption{Ellipticity histograms for galaxies with $24.0 \le
$ V $\le 26.5$ in the 0957+561 field (solid line) and one of the simulations
(dashed line) $24.0 \le $ V $\le 26.5$. The distributions are similar, and the
mean ellipticities are $\epsilon = 0.265$ and 0.275,
respectively. \label{ellip}}
\end{figure}

Fig. \ref{lumfunc} shows the observed galaxy counts of one of the simulated
data frames superimposed on the 0957+561 galaxy counts. Assuming the input
galaxy counts are correct, the completeness curves are in good agreement,
giving us confidence that the size-magnitude relationship is close to
correct. Fig. \ref{ellip} shows ellipticity histograms for the data and one of
the simulations. These are a close match, and have similar mean ellipticities
of $\epsilon = 0.265 and 0.275$, respectively. It is important that the
simulated and real galaxies are well matched in both ellipticity and size since
the response of a galaxy under shear (the ``shear polarizability'' - see Kaiser
et al. 1995) is dependent on these two parameters. The results of the Monte
Carlo simulations are discussed in the next section.

\section{Discussion} \label{discuss}

\subsection{2-d Mass Maps} \label{2d}

The 2-d, KS mass map for our data is shown in Fig. \ref{contour} superposed on
the V-band image of the field.  This reconstruction used 1651 galaxies with
$24.0 \le V \le 26.5$. The detection of the central peak is significant at
around the $4.5\sigma$ level based on the scatter in the simulations.

\begin{figure}
\plotone{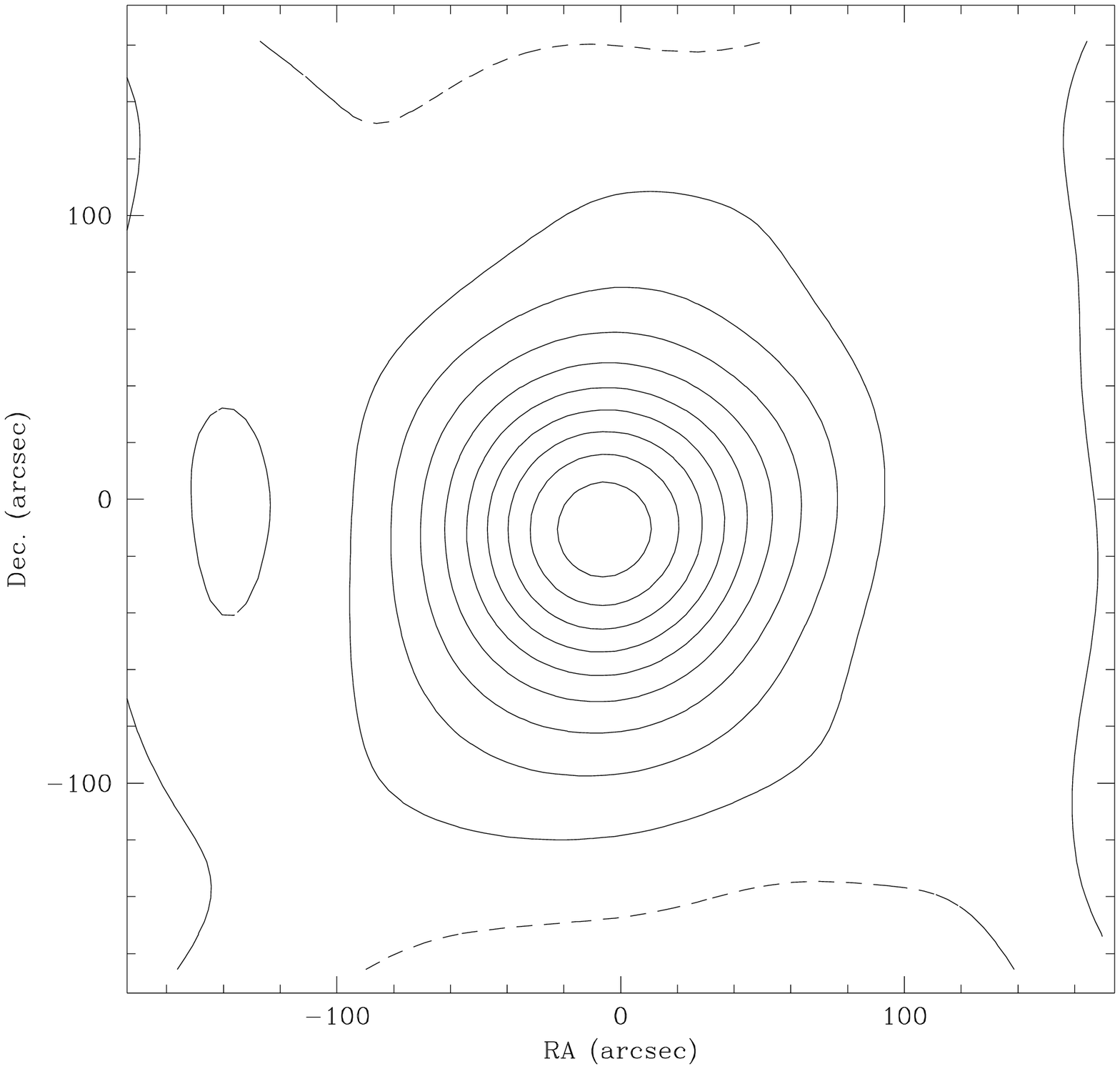}
\caption{Mean mass map for 102 simulations of a spherical
cluster. The smoothing scale is $s=30\arcsec$. The contours have steps of
approximately $0.5\sigma$ of an individual image. Edge effects are apparent as
well as a small vertical elongation. \label{consim}} 
\end{figure}

The first thing the simulations can reveal are systematic errors in the 2-d
mass map due to the elongated PSF and the finite size of the image. In
Fig. \ref{consim} we show the average mass map from 102 Monte Carlo simulations
of a spherical cluster which fits our data well (\S \ref{profile}).  There is
some evidence for small spurious features: a slight elongation in the
north-south direction and $1\sigma$ negative features along all four edges.
These systematics are what is expected from the KS algorithm (Schneider 1995),
and become more serious as the distance from the image center increases.
Therefore, in order to quantify measurements made with the mass map we will use
the simulations as a comparison.

The isodensity contours in Fig. \ref{contour} are not identical to the mean of
the simulations. Is this a significant departure from circular symmetry or just
the noise in the KS reconstruction? The issue of substructure is an important
one from the perspective of H$_0$ determination. There is spectroscopic
evidence for the existence of a poor cluster at z=0.5 (Garrett et al. 1992,
Angonin-Willaime et al. 1994) centered about 81\arcsec\ west and 43\arcsec\
north of G1. As pointed out in Bernstein et al. (1993), if sufficiently
massive, this cluster may be an important constituent of the lens.

\begin{figure}
\plotone{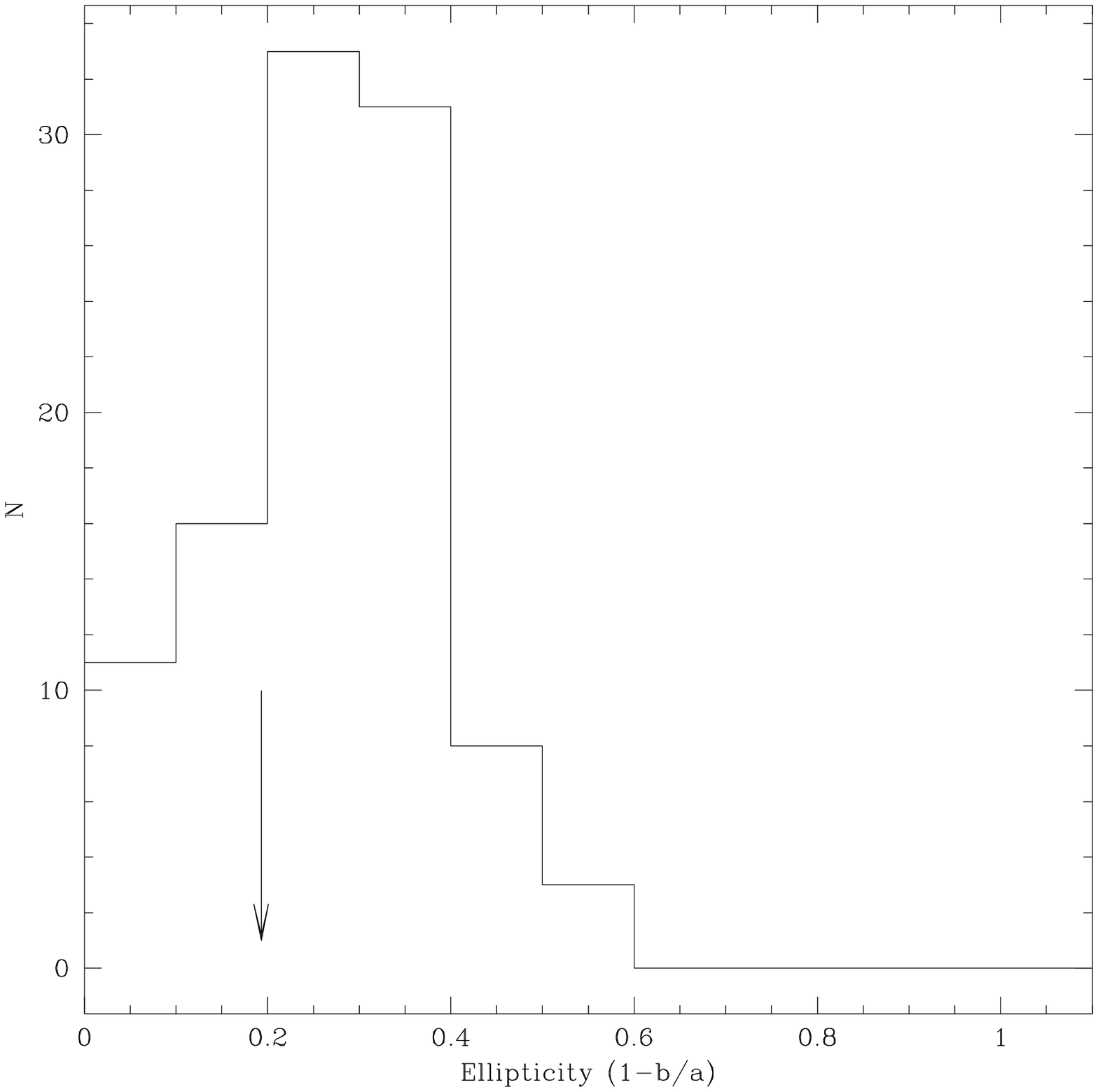}
\caption{Ellipticy histograms of the simulated mass
maps. The input models have $\epsilon = 0$ The arrow indicates the measured
value for the 0957+561 field. \label{quad}}
\end{figure}

In order to test the significance of the deviations from circularity in the KS
mass reconstruction (Fig. \ref{contour}), we compute its ellipticity from the
quadrupole moments of the portions of the map with $\kappa >0$.  Fig.
\ref{quad} shows this ellipticity value along with the distribution of
ellipticities measured in an identical manner from the Monte-Carlo simulations,
which all contain perfectly circular mass distributions. The 0957+561 map is a
bit rounder than the mean of the simulations but well within the normal
range. Therefore we conclude that we have not detected substructure in our
cluster. It should be noted that we have used a smoothing scale of $s =
30\arcsec$ and substructure finer than this would be washed out.  Probing
higher resolution substructure needs a higher density of background galaxies,
requiring deeper images with better seeing.

Also of interest for H$_\circ$ determination is the accuracy of the mass
centroid measurement. The mean of the centers of the 102 simulations show no
systematic deviations from the input center. The 1$\sigma$ uncertainty in the
position of the centers is 15\arcsec.
From Fig. \ref{contour}, the center of mass in the 0957+561 field is located
$\sim22\arcsec$ from the galaxy G1 (18\arcsec\ east 13\arcsec\ north of G1).
consistent with the position of G1 at the $1.5\sigma$ level. The direction of
the offset appears to be consistent with that found by Dahle et al. (1994)
using 123 galaxies (compared to 1651 in this study) although our offset is
larger. These two measurements are not completely independent since many of the
arclets near the QSO will be common to both studies.  Fig. \ref{galdens} is a
contour plot of galaxy number density for $V<24$. There is a $3\sigma$ peak
offset from G1 in the same direction as the mass peak lending support to the
idea that G1 is not at the center of mass.  In order to avoid biasing our
result, we have adopted G1 as the mass center for our azimuthally averaged mass
density profile described below.

\begin{figure}
\caption{Contours of galaxy surface number density for
$V<24$ in steps of $0.5\sigma$ about the mean. A total of 302 galaxies were
used and a Gaussian smoothing with $\sigma = 30\arcsec$ has been applied.
\label{galdens}} 
\end{figure}

\subsection{Mass Profile} \label{profile}

The azimuthally averaged mass density profile centered on the galaxy G1 as
derived from the weak lensing approximation (Eqn. \ref{denseqn}) is shown in
the upper panel of Fig. \ref{compdens}. The individual points shown in this
plot are not independent since values at large radii contain a subset of the
galaxies used to determine quantities at smaller radii.

Comparing our surface density profile to a variety of spherical cluster
simulations with isothermal plus core profiles we find good agreement between
data and simulations for:

\begin{equation} \label{iso}
\Sigma(r) = \Sigma_0[1+(r/r_c)^2]^{-1/2},
\end{equation}

\noindent with $r_c = 5\arcsec = 17 h^{-1}$ kpc and $\Sigma_0 = 3600 \pm 1100
h$ M$_\odot$ pc$^{-2}$. There is some marginal evidence that the mass profile
falls off faster than isothermal for $r > 400$ kpc $h^{-1}$ however the
uncertainties are large at these radii and a bigger field is required to test
this hypothesis.

As can be seen in the upper panel of Fig. \ref{compdens} the mean surface mass
density of the simulated clusters determined under the weak lensing
approximation is slightly steeper than the input surface mass density. This
leads to an approximately 20\% overestimate of the mass at $r = 10\arcsec$.
This can be attributed to a breakdown in the weak lensing
approximation. Because we know the true density profile for these simulations
we can test this by removing the weak lensing assumption and seeing if the
agreement improves. This amounts to modifying Eqn. \ref{denseqn} to the
following form:

\begin{equation} \label{nonweakeqn}
\overline\kappa(r \le r_i) - \overline\kappa(r_i \le r \le r_o) =
{r_o^2 \over M}\sum_{r_i \le r \le
r_o}{[1-\kappa(\vec{R})][1-\sqrt{1-D_i(\vec{R})^2}]
\over D_i(\vec{R})(\Delta{x_i}^2 + \Delta{y_i}^2)}
\end{equation}

\noindent
The value of $\kappa(\vec{R})$ for each galaxy is calculated assuming the mean
$\Sigma_{crit}$ of the background galaxy sample. From the Monte-Carlo
simulations we find $\Sigma_{crit}=5650 h$ M$_\odot$ pc$^{-2}$.  The lower
panel of Fig. \ref{compdens} shows the results of the modified mass
inversion. The agreement between the input and inferred mass density is
excellent, with no sign of systematic deviations. The value of the recovery
factor is $<C> = 0.36 \pm 0.02$, very similar to the constant shear HDF result.
Therefore, for subcritical lenses, one can get accurate density estimates
either by fitting the data directly to the results of the simulations (which
should suffer from the same systematics) or, non-parametrically, by adopting an
iterative approach where the first step assumes the weak lensing approximation
and subsequent steps use the previous value of the (smoothed) density profile
to estimate the $\gamma_T$ and $(1 - \kappa)$ terms in
Eqn. \ref{nonweakeqn}. We have adopted the former technique and show the
corrected profile in the lower panel of Fig. \ref{compdens}.

\begin{figure}
\plotone{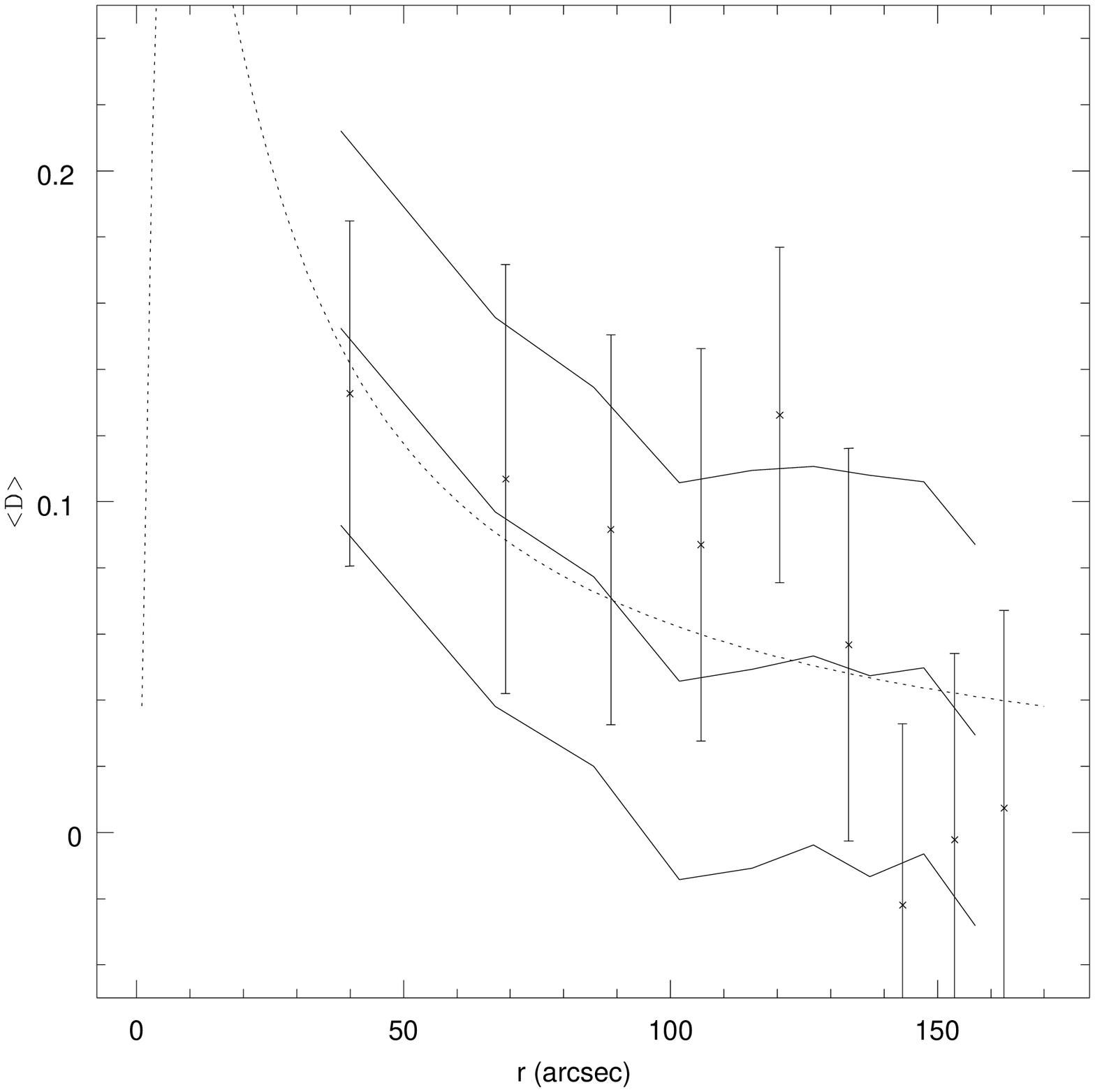}
\caption{A plot of tangential shear vs projected radius in
radial bins containing 145 galaxies each ($10\arcsec\ \le r \le 168\arcsec$,
$24.0 \le V \le 26.5$), centered on G1. The points are from the data and the
solid lines are the mean and $1\sigma$ upper and lower error bands from 102
Monte-Carlo simulations. The dashed line is the input model for the simulations
(Eqn. \protect\ref{iso}) and has $r_c = 5\arcsec$ and $\Sigma_0 = 3600
\Sigma_{crit}$. The dashed line is {\it not} averaged over the bins.
\label{shear}}
\end{figure}

Fig. \ref{shear} shows a plot of calibrated $<D(r)>$ in radial bins around
G1. These points, though noisy, are independent.  The reduced $\chi^2$ between
the points and model line in Fig. \ref{shear} is 0.62. Due to the large error
bars on the points it is not possible to provide stong constraints on the shape
of the profile. The value of $<D>$ for the entire radial range $10\arcsec\ \le
r \le 168\arcsec$ is $0.064 \pm 0.019$ (s/n = 3.3).

The Monte-Carlo simulations provide a good check on our error analysis. In
Fig. \ref{compdens} we show the surface density and uncertainty derived from
the data along with the mean and scatter of the simulations. The two sets of
error estimates are in good agreement. The s/n of the density estimates
decreases at the smallest radii due to sensitivity to the presence of a small
number of radially or tangentially aligned galaxies. This results because of
${1 \over \Delta{x^2} + \Delta{y^2}}$ in the denominator of
Eqn. \ref{denseqn}. For example, by removing the four most extreme galaxies,
the apparent s/n increases from around 3 to 4.

As mentioned, the best fit model to the data has $r_c = 5\arcsec\ = 17 h^{-1}$
kpc. However, this is quite dependent on the innermost point in
Fig. \ref{compdens} which, as just described, is very sensitive to a small
number of galaxies. Therefore, it is possible that the core radius could be
substantially in error. At the low end, it is impossible to rule out $r_c =
0$. At the high end, lowering the innermost point $1\sigma$ yields $r_c =
10\arcsec\ = 33 h^{-1}$ kpc.

One systematic that we do not account for in our Monte-Carlo simulations is
contamination of our field galaxy sample by cluster galaxies. If cluster
galaxies have been included in our mass reconstruction then we will have
underestimated the mass density. If the cluster galaxies are concentrated
towards the center of the cluster (as one would expect) then we would also be
underestimating the steepness of the mass profile. Because of the difficulty in
distinguishing possible cluster galaxies from the field galaxy population we
have not attempted to correct for contamination. However, a surface number
density map of all galaxies used in the mass reconstruction ($24 < V < 26.5$)
reveals no significant concentrations centered near G1 (or any significant
density fluctuations in the field), so we conclude that any cluster galaxy
contamination is small.

The surface mass density estimates discussed in this section are determined
from measurments of density {\it contrasts} between innner regions and control
annuli (Eqn. \ref{denseqn}). Therefore an estimate of the {\it total} surface
density requires that we estimate the mass densities of the control annuli.
This was implicit in our assumption of the functional form in Eqn \ref{iso} for
the cluster mass distribution.  Because this estimate is based on an
extrapolation of our measurements to a region which we cannot measure, it is
worth mentioning the size of the correction. For the innermost point ($r =
10\arcsec$), the mean density of the control annulus is around 7\% of the
interior mean density if the density profile follows Eqn \ref{iso}. This rises
to about 50\% for our outermost point. There is some marginal indication from
the density profile that the mass density falls off more sharply than our
model. If this is true then the mass densities of the control annuli are
overestimated and the mass densities in this paper will be
overestimated. Alternatively, if the mass densities in the control annuli are
underestimated (perhaps there is a flat component to the cluster mass density),
then the mass densities are underestimated.

\section{Other Mass Determinations} \label{other}

Previous attempts to measure the cluster mass in the 0957+561 field include a
radial velocity study and an X-ray study.

Garrett et al (1992) and Angonin-Willaime et al. (1994) measured redshifts for
21 probable cluster members, obtaining a velocity dispersion of $\sigma = 715
\pm 130$ km s$^{-1}$. If one assumes a singular isothermal mass distribution
and an isotropic phase space distribution function, the implied mass contained
within a projected radius of 1 Mpc is M($r<1$ Mpc) $=3.7^{+1.5}_{-1.2} \times
10^{14}$ M$_\odot$. Of course uncertainties in the distribution function and
problems with field galaxy contamination imply a much higher uncertainty. In
any case, extrapolating our model with $\Sigma_0 = 3600 \pm 1100 h$ M$_\odot$
pc$^{-2}$ and $r_c = 17 h^{-1}$ kpc implies a mass of M($r<1$ Mpc) $= 3.9 \pm
1.2 \times 10^{14}$ M$_\odot$, consistent with the dynamical mass estimate.

Based on ROSAT PSPC and HRI data of the 0957+561 field, Chartas et al. (1995)
place upper limits on the cluster temperature and mass. They obtained a
$3\sigma$ upper limit on the cluster mass within a projected radius of 1 Mpc of
G1 of $1.5 \times 10^{14}$ M$_\odot$, $2\sigma$ lower than the lensing mass
estimate.


\section{Conclusion} \label{conclusion}

The double QSO Q0957+561 was the first multiply gravitationally lensed system
discovered (Walsh et al. 1979). It is currently the only lensed system with a
firm time-delay. Extensive modeling has been carried out for this system and
the conclusion is that in order to constrain H$_0$ the mass distribution in the
lens must be better understood (Grogin \& Narayan 1996 and references
therein). The lens consists of the galaxy G1 and the cluster to which G1
belongs.

In this paper we have determined the mass distribution in the z=0.355 cluster
in the field 0957+561 by studying the distorted appearance of faint background
galaxies with $24.0 \le V \le 26.5$. We derived the 2-d mass distribution which
appears to be consistent with a spherical cluster given our smoothing scale of
$s = 30\arcsec$. The mass centroid is located 22\arcsec\ from the dominant
cluster galaxy G1, but is consistent with the galaxy position at the
$1.5\sigma$ level. A density map of galaxies with $V<24.0$ has a peak, offset
from G1 in the same direction as the mass map strengthening the argument that
G1 is offset from the center of mass.

We have calculated the azimuthally averaged mass profile centered on G1
out to $400 h^{-1}$ kpc. It is well fit by an isothermal model possessing a
core of $r_c = 5\arcsec\ = 17 h^{-1}$ kpc. 

In terms of total mass, good agreement is seen with previous kinematic
measurements under the assumption of isotropy. Our gravitational lensing mass,
however, is substantially higher than upper limits based on X-ray data.

We will discuss the implications for H$_0$ in a companion paper.


\acknowledgments

Support for this work was provided by NASA through grant \# HF-01069.01-94A
from the Space Telescope Science Institute, which is operated by the
Association of Universities for Research in Astronomy Inc., under NASA contract
NAS5-26555.  George Rhee acknowledges travel support from the UNLV Physics
department Bigelow fund.  Thanks to: Ian Smail for providing his V-band galaxy
counts, Brian McLeod for assistance in reproducing his galaxy evolution
models and Peter Schneider for pointing out an error in an earlier version of
this paper. 

{}


\begin{thebibliography}{}

\bibitem[]{} Angonin-Willaime, M.-C., Soucail, G. \& Vanderriest, C. 1994,
A\&A, 291, 411.

\bibitem[]{} Bernstein, G. M., Tyson, J. A., \& Kochanek,
C. S. 1993, AJ, 105, 816.

\bibitem[]{} Bruzual, A. G., \& Charlot, S. 1993, ApJ, 405, 538.

\bibitem[]{} Burstein, D. \& Heiles, C. 1982, AJ, 87, 1165.

\bibitem[]{} Chartas, G., Falco, E. Forman, W., Jones, C. Schild, R. \&
Shapiro, I., ApJ, 445, 140.

\bibitem[]{} Christian, C. A. et al. 1985, PASP 97 363.

\bibitem[]{} Dahle, H., Maddox, S. J., \& Lilje, P. B. 1994, ApJL, L79.

\bibitem[]{} Fahlman, G.G., Kaiser, N., Squires, G. \& Woods, D.
1994, ApJ, 436, 56.

\bibitem[]{} Fukugita, M., Shimasaku, K. \& Ichikawa, T. 1995, PASP, 107, 945.

\bibitem[]{} Garrett, M. A., Walsh, D. \& Carswell, R. F. 1992, MNRAS, 254,
27p.

\bibitem[]{} Grogin, N. A. \& Narayan, R. 1996, ApJ, 464, 92.

\bibitem[]{} Im, M., Casertano, S. Griffiths, R. E., Ratnatunga,
K. U. \& Tyson, J. A. 1995, ApJ, 441 494.

\bibitem[]{} Kaiser, N. \& Squires, G. 1993, ApJ, 404, 441

\bibitem[]{} Kaiser, N., Squires, G. \& Broadhurst, T. 1995, ApJ, 449, 460.

\bibitem[]{} King, C. \& Ellis, R. 1985, ApJ, 288, 456.

\bibitem[]{} Kundic et al. 1996, preprint.

\bibitem[]{} McLeod, B. A., Rieke, M. J. 1995, ApJ, 454, 611.

\bibitem[]{} Miralda-Escud\'e, J. 1991, ApJ, 370, 1.

\bibitem[]{} Miralda-Escud\'e, J. 1995, in ``IAU 173: Astrophysical
Applications of Gravitational Lensing'', eds. C. S. Kochanek \& J. N. Hewitt, 
(Kluwer), p. 131.

\bibitem[]{} Refsdal, S. 1964, MNRAS, 128, 307

\bibitem[]{} Rhee, G. 1991, Nature, 350, 211

\bibitem[]{} Schneider, P. 1995, A\&A, 302, 639.

\bibitem[]{} Schneider, P. \& Seitz, C. 1995, A\&A, 294, 411.

\bibitem[]{} Schild, R, \& Thomson, D. J. 1995, AJ, 109, 1970.

\bibitem[]{} Seitz, C. \& Schneider, P. 1995, A\&A, 297, 287.

\bibitem[]{} Smail, I, Hogg, D. W., Yan, L. \& Cohen, G. 1995,
ApJL, L105.

\bibitem[]{} Squires, G, et al 1996, ApJ, 461, 572.

\bibitem[]{} Thomson, D. J., \& Schild, R, 1994, in Proceedings of the
International Conference on Applications of Time Series Analysis in Astronomy
and Meteorology (in press).

\bibitem[]{} Tyson, J. A. \& Fischer, P. 1995, ApJL, L55.

\bibitem[]{} Tyson, J.A., Valdes, F. \& Wenk, R.A. 1990, ApJ,
349, L1.

\bibitem[]{} Walsh, D., Carswell, R. F., Weymann, R. J., Nature,
279, 381.

\bibitem[]{} Williams, R. E. et al. 1995, BAAS, 187.0903.

\bibitem[]{} Young, P., Gunn, J. E., Kristian, J., Oke, J. B., \& Westphal,
J. 1981, ApJ, 244, 736.

\end{thebibliography}
\end{document}